\documentclass[journal,twoside,twocolumn,a4paper]{IEEEtran}

\usepackage{amsmath,amssymb}
\usepackage[dvips,colorlinks=true,bookmarks=false,citecolor=blue,urlcolor=blue]{hyperref} 
\usepackage{graphicx,wrapfig}
\usepackage{psfrag}
\usepackage{cite}
\usepackage{balance}

\newcommand{\Real}{\mathbb{R}}
\newcommand{\tnr}[1]{{\textnormal{#1}}}
\newcommand{\E}{\mathbb{E}}
\newcommand{\mc}[1]{\mathcal{#1}}

\newcommand{\BERin}{\tnr{BER}_{\tnr{pre}}}
\newcommand{\BERout}{\tnr{BER}_{\tnr{pos}}}
\newcommand{\Cfs}{\mathcal{C}_{4,16}}
\newcommand{\Cft}{\mathcal{C}_{4,256}}
\newcommand{\Cff}{\mathcal{C}_{4,4096}}

\newcommand{\bB}{\boldsymbol{B}}
\newcommand{\bU}{\boldsymbol{U}}\newcommand{\bu}{\boldsymbol{u}}
\newcommand{\bX}{\boldsymbol{X}}\newcommand{\bx}{\boldsymbol{x}}
\newcommand{\bY}{\boldsymbol{Y}}\newcommand{\by}{\boldsymbol{y}}
\newcommand{\bZ}{\boldsymbol{Z}}
\newcommand{\bs}{\boldsymbol{s}}

\newcommand{\argmax}{\mathop{\mathrm{argmax}}}

\newcommand{\mcC}{\mathcal{C}}

\newcommand{\mcS}{\mathcal{S}}
\newcommand{\mcSkb}{\mathcal{S}_{k,b}}
\newcommand{\mcSko}{\mathcal{S}_{k,1}}
\newcommand{\mcSkz}{\mathcal{S}_{k,0}}
\newcommand{\set}[1]{\{#1\}}
\newcommand{\ie}{i.e.,~}

\newcommand{\cd}{\cdot}

\newcommand{\ld}{\ldots}

\newcommand{\Eb}{E_\tr{b}}
\newcommand{\No}{N_{0}}
\newcommand{\Es}{E_\tr{s}}
\newcommand{\Ns}{N_\tr{s}}
\newcommand{\Nb}{N_\tr{b}}
\newcommand{\tr}[1]{\mathrm{#1}}

\newcommand{\un}[1]{\underline{#1}}
\newcommand{\figref}[1]{Fig.~\ref{#1}}
\newcommand{\figsref}[2]{Figs.~\ref{#1} and \ref{#2}}

\newcommand{\secref}[1]{Sec.~\ref{#1}}

\newcommand{\Rc}{R_\tr{c}}
\newcommand{\R}{\eta}

\newcommand{\overlay}[3]{\makebox[0mm][l]{\hspace*{#1}\raisebox{#2}[0ex][0ex]{#3}}}

\usepackage[ddmmyyyy]{datetime}
\newcommand{\header}{Preprint, \today, \currenttime}
\title{Four-Dimensional Coded Modulation with Bit-wise Decoders for Future Optical Communications}
\author{Alex Alvarado and Erik Agrell
\thanks{Research supported by Engineering and Physical Sciences Research Council (EPSRC) project UNLOC (EP/J017582/1), United Kingdom, by the Swedish Research Council (VR) under grant no.~2012-5280. Parts of this paper were presented at the 2014 Optical Fiber Communication Conference (OFC), San Francisco, CA, Mar. 2014
and at OFC 2015, Los Angeles, CA, Mar. 2014.}
\thanks{A.~Alvarado is with the Optical Networks Group, Department of Electronic and Electrical Engineering, University College London, London WC1E~7JE, United Kingdom (email: alex.alvarado@ieee.org).}
\thanks{E.~Agrell is with the Department of Signals and Systems, Chalmers University of Technology, SE-41296 Gothenburg, Sweden (email: 
agrell@chalmers.se).}
}
\begin{document}
\maketitle

\begin{abstract}
Coded modulation (CM) is the combination of forward error correction (FEC) and multilevel constellations. Coherent optical communication systems result in a four-dimensional ($4$D) signal space, which naturally leads to $4$D-CM transceivers. A practically attractive design paradigm is to use a bit-wise decoder, where the detection process is (suboptimally) separated into two steps: soft-decision demapping followed by binary decoding. In this paper, bit-wise decoders are studied from an information-theoretic viewpoint. $4$D constellations with up to 4096 constellation points are considered. Metrics to predict the post-FEC bit-error rate (BER) of bit-wise decoders are analyzed. The mutual information is shown to fail at predicting the post-FEC BER of bit-wise decoders and the so-called generalized mutual information is shown to be a much more robust metric. For the suboptimal scheme under consideration, it is also shown that constellations that transmit and receive information in each polarization and quadrature independently (e.g., PM-QPSK, PM-16QAM, and PM-64QAM) outperform the best $4$D constellations designed for uncoded transmission. Theoretical gains are as high as 4~dB, which are then validated via numerical simulations of low-density parity check codes.
\end{abstract}


\section{Introduction and Motivation}\label{Sec:Introduction}

In coherent fiber-optic communication systems, both quadratures and both polarizations of the electromagnetic field are used. This naturally results in a four-dimensional ($4$D) signal space. To meet the demands for spectral efficiency, multiple bits should be encapsulated in each constellation symbol, resulting in multilevel $4$D constellations. To combat the decreased sensitivity caused by multilevel modulation, forward error correction (FEC) is used. The combination of FEC and multilevel constellations is known as \emph{coded modulation} (CM). 

The most popular alternatives for CM are trellis-coded modulation (TCM) \cite{Ungerboeck82}, multilevel coding (MLC) \cite{Imai77}, and bit-interleaved coded modulation (BICM) \cite{Zehavi92,Caire98,Fabregas08_Book}. TCM has been considered for optical communications in \cite{Benedetto95d,Bulow04,Zhao06,Kumar07,Magarini10} and MLC in \cite{Djordjevic2006_JLT,Gong10,Beygi10b,Smith12b,Farhoudi14}. Regardless of the paradigm used at the transmitter (see \cite[Fig.~3]{Beygi14} for a schematic comparison), the optimum receiver structure is the maximum likelihood (ML) decoder. The ML decoder finds the most likely transmitted sequence, where the maximization is over all possible \emph{coded} sequences. The ML solution is in general impractical\footnote{A notable exception is TCM, where the FEC encoder is a convolutional encoder and the resulting CM code has a trellis structure, which allows an ML decoder based on the Viterbi algorithm to be implemented.}, and thus, suboptimal alternatives are preferred. One pragmatic and popular approach is BICM, which we study in this paper.

The key feature of BICM is a \emph{suboptimal} decoder that operates on bits rather than on symbols. We refer to this receiver structure as a \emph{bit-wise} (BW) decoder. In a BW decoder, the detection process is decoupled: first soft information on the bits (logarithmic likelihood ratios, LLRs) is calculated in a \emph{demapper} and then a soft-decision FEC (SD-FEC) decoder is used. BW decoders are very flexible, where the flexibility is due to the use of off-the-shelf \emph{binary} encoders and decoders. In the context of optical communications, a BW decoder for binary modulation and low-density parity check (LDPC) codes was studied in \cite{Djordjevic2006_JSQE}, where a finite-state machine and a histogram-based estimation of the channel was used to compute LLRs. A BW decoder with multilevel modulation and LDPC codes was considered in \cite{Bulow2009}. An LDPC-based BW decoder with a $24$-dimensional constellation was experimentally demonstrated in \cite{Millar14}. Optimized mappings between code bits and constellation symbols for protograph-based LDPC codes were recently presented in \cite{Hager14a}.

To improve upon simple BW decoders, iterations between the binary FEC decoder and demapper can be included. In such a configuration, the FEC decoder and demapper iteratively exchange information on the code bits. This is usually known as BICM with iterative demapping (BICM-ID). \mbox{BICM-ID} for optical communications has been studied in \cite{Djordjevic2007_JLT,Batshon2009_JLT, Bulow14}, \cite[Sec.~3]{Bulow2011b}, \cite[Sec.~3]{Bulow2011}, \cite[Sec.~4]{Schmalen14}. BICM-ID offers remarkable improvements with demapper iterations. These gains are typically obtained by custom-tailoring the constellation and its binary labeling to the channel and the encoder--decoder pair as well as the iteration scheduling \cite{Schmalen14}. In BICM-ID, iterations between the decoder and demapper are added to a possibly already iterative FEC decoder and to keep the number of iterations low, one can trade FEC decoder iterations for demapper iterations. However, this leads to nontrivial designs which reduce flexibility. On the positive side, BICM-ID is expected to perform very close to an ML sequence detector, and thus, to outperform BICM. To the best of our knowledge, no exact complexity-performance tradeoff analyses providing a clear-cut answer about BICM vs. BICM-ID exist. In this paper, we focus on BICM because of its simplicity and flexibility.

CM transceivers are typically based on quadrature amplitude modulation (QAM) or phase shift keying (PSK). Traditional constellations include polarization-multiplexed (PM) quadrature phase-shift keying (PM-QPSK)\footnote{Also known in the literature as dual-polarization QPSK (DP-QPSK) and polarization-division-multiplexed QPSK (PDM-QPSK).}, PM-16QAM, and PM-64QAM. However, recent years have seen an increased interest in
formats that use the available four dimensions more efficiently than by pure multiplexing.
Polarization-switched QPSK (PS-QPSK) was shown in \cite[Fig.~1]{Karlsson2009} to be the most power-efficient 8-ary $4$D constellation. Power efficiency should here be understood as the energy per bit for a given minimum Euclidean distance between constellation points. This is the classical sphere packing problem, which has been used to optimize constellation formats for uncoded transmission since the 1970's \cite{Simon73,Foschini74,Agrell-codes}. It arises when minimizing either the pre-FEC bit error rate (BER) or the symbol error rate for the additive white Gaussian noise (AWGN) channel at asymptotically high signal-to-noise ratio (SNR) \cite{Agrell2009_JLT, Agrell-codes}, \cite[Sec.~5.1]{Benedetto99_Book}. $4$D constellations optimized in this sense were compared in \cite{Agrell2009_JLT}. Spherically shaped $4$D constellations based on the $D_{4}$ lattice were studied in, e.g., \cite{Karlsson12OFC, Bulow14}. Somewhat less power efficient, but easier to implement, are the cubically shaped constellations based on $D_4$, called set-partitioning QAM \cite{Coelho11,Karlsson12OFC}.
Other irregular constellations include the amplitude phase-shift keying constellation optimized for channels with strong nonlinear phase noise in \cite{Pfau11,Beygi11,Hager13}.

Of particular interest for this paper is the constellation $\Cfs$ introduced in \cite{Karlsson2010}, which is the most power efficient $16$-ary $4$D constellation known. Another constellation we will study in this paper is subset-optimized PM-QPSK (SO-PM-QPSK) introduced in \cite{Sjodin2013_CL} as an alternative to $\Cfs$ with lower complexity. In terms of power efficiency, $\Cfs$ and SO-PM-QPSK offer asymptotic gains over PM-QPSK of $1.11$~dB and $0.44$~dB, respectively. The asymptotic gains offered by $\Cfs$ have been experimentally demonstrated in \cite{Karout2013_OFC, Bulow2013}. We also consider the power-efficient $4$D constellations $\Cft$ \cite[Table.~IV]{Welti74}, \cite[Table~I]{Conway83} and $\Cff$ \cite{Agrell-codes}, which are, respectively, the best known $256$-ary and $4096$-ary constellations.

The performance of a BW decoder based on hard decisions (HDs) can be accurately characterized by the pre-FEC BER. In this paper, we study SD-FEC, i.e., when LLRs are passed to the soft-input FEC decoder, and thus, we question the optimality of constellations designed in terms of pre-FEC BER. Furthermore, we show that a different metric is more relevant for capacity-approaching SD-FEC encoder--decoder pairs: the so-called \emph{generalized mutual information} (GMI).

Achievable rates provide an upper bound on the number of bits per symbol that can be reliably transmitted through the channel. From an information-theoretic point of view, a BW decoder does not implement the ML rule, and thus, a penalty in terms of achievable rates is expected. While the \emph{mutual information} (MI) is the largest achievable rate for any receiver, for a BW decoder, this quantity is replaced by the GMI \cite[Sec.~3]{Fabregas08_Book}\footnote{The term GMI was coined by Martinez \emph{et al.} in \cite{Martinez09}, where the BW decoder was recognized as a mismatched decoder. The GMI is known in  the literature under different names such as ``parallel decoding capacity'', ``receiver constrained capacity'', and ``BICM capacity.''} Although the MI and the GMI coincide when the SNR tends to infinity, for any nontrivial case, the MI is strictly larger than the GMI for any finite SNR. This penalty, which depends on the constellation and its binary labeling, can be very large \cite[Fig.~4]{Caire98}. The MI has been considered as the figure of merit for optical communications in \cite{Essiambre10,Karlsson2010,Goebel11b,Smith12b,Karout2013_OFC,Poggiolini14a,Fehenberger14a,Khodakarami13}. To the best of our knowledge, however, the GMI has been considered in optical communications only in \cite{Bulow2011b}.

One problem often overlooked when designing $4$D-CM with a BW decoder is the problem of choosing an appropriate binary labeling for the constellation. Finding good labelings based on brute force approaches quickly fails, as the number of binary labelings grows factorially with the constellation size. For example, for the relatively simple case of $16$ constellation points, there are about $2\cd10^{13}$ different binary labelings. When regular constellations (QAM, PSK, etc.) are considered, a Gray code\footnote{In fact, Gray codes are not unique, and the one often used is the so-called binary reflected Gray code (BRGC) introduced in 1953 \cite{Gray53}.} is typically used, as Gray codes have been proven to be asymptotically optimum in terms of pre-FEC BER \cite{Agrell07}. This conclusion holds only in the regime of asymptotically large SNR and only for the AWGN channel. The problem is considerably more difficult when the GMI is the cost function. Although results in the asymptotic regimes exist (see \cite{Martinez08b, Alvarado10c, Stierstorfer09a,Agrell10b,Agrell12c} and \cite{Alvarado12b} for low and high SNR, respectively), finding the optimal binary labeling in terms of GMI for a finite SNR remains as an open research problem.

In this paper, achievable rates for $4$D constellations with a BW decoder in the context of future generation coherent optical communication systems are studied. It is shown that the GMI is the correct metric to predict the post-FEC BER for a BW decoder. It is also shown that constellations that are good for uncoded systems are also good in terms of MI if the SNR is sufficiently high. These constellations, however, are not  the best choice for coded systems based on a BW decoder. Numerical results based on LDPC codes confirm the theoretical analysis. 

The remainder of this paper is organized as follows. In \secref{Sec:Model}, the system model is introduced and achievable rates are reviewed. Post-FEC BER prediction based on the GMI is studied in \secref{Sec:NumResults:GMI} and numerical results on achievable rates are shown in \secref{Sec:NumResults}. Conclusions are drawn in \secref{Sec:Conclusions}. 

\begin{figure*}[t]
\begin{center}
\newcommand{\scale}{0.84}
\psfrag{bi}[br][Br][\scale]{$\un{\bU}$}
\psfrag{CHE1}[ll][cc][\scale]{\overlay{-4.3mm}{1mm}{Binary}}
\psfrag{CHE2}[ll][cc][\scale]{\overlay{-2.8mm}{2mm}{FEC}\overlay{-5.5mm}{-2mm}{Encoder}}
\psfrag{bc1}[cc][cc][\scale]{$\un{B}_{1}$}
\psfrag{bcm}[cc][cc][\scale]{$\un{B}_{m}$}
\psfrag{MOD1}[cc][cc][\scale]{Memoryless}
\psfrag{MOD2}[cc][cc][\scale]{Mapper}
\psfrag{ENCODER}[Br][Br][\scale]{CM Encoder}
\psfrag{DECODER}[Bl][Bl][\scale]{CM Decoder}
\psfrag{COMM}[Bc][Bc][\scale]{Communication}
\psfrag{CHANNEL}[Bc][Bc][\scale]{Channel}
\psfrag{x}[bc][Bc][\scale]{$\un{\bX}$}
\psfrag{ddd}[cc][cc][\scale]{$\vdots$}
\psfrag{y}[bc][Bc][\scale]{$\un{\bY}$}
\psfrag{DEC1}[cc][cc][\scale]{ML or BW}
\psfrag{DEC2}[cc][cc][\scale]{Decoder}
\psfrag{bih}[bl][Bl][\scale]{$\un{\hat{\bU}}$}
\includegraphics[width=0.80\linewidth]{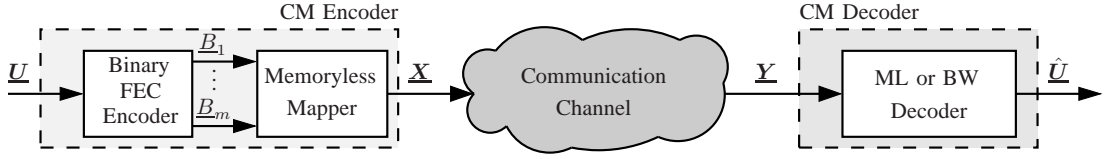}
\end{center}
\caption{CM structure under consideration. The CM encoder is a concatenation of a rate-$\Rc$ binary FEC encoder and a memoryless mapper. The CM decoder is either an ML decoder or a BW decoder (see \figref{receivers}).}
\label{model}
\end{figure*}

\section{System Model and Achievable Rates}\label{Sec:Model}

In \figref{model}, a generic structure of the CM transceiver we study in this paper is shown. At the transmitter, a rate-$\Rc$ binary FEC encoder encodes a binary input sequence $\un{\bU}$ into $m$ binary sequences $\un{B}_1, \ldots, \un{B}_m$, where $\un{B}_k = [B_{1,k},\ldots,B_{\Ns,k}]$ for $k=1,\ldots,m$ and $\Ns$ is the symbol block length.\footnote{Throughout this paper, vectors are denoted by boldface letters $\bx$, sequences of vectors by underlined boldface letters $\un{\bx}$, and sets by calligraphic letters $\mc{X}$. Random variables, vectors, and sequences are denoted by uppercase letters and their outcomes by the same letter in lowercase. Probability density functions and conditional probability density functions are denoted by $f_{\bY}(\by)$ and $f_{\bY|\bX}(\by|\bx)$, respectively. Expectations are denoted by $\E[\cd]$.}. A memoryless mapper then maps $\un{B}_1, \ldots, \un{B}_m$ into a sequence of symbols $\un{\bX}=[\bX_{1},\bX_{2},\ld,\bX_{\Ns}]$, one symbol at a time. After transmission over the physical channel, the received symbols $\un{\bY}=[\bY_{1},\bY_{2},\ld,\bY_{\Ns}]$ are processed by the CM decoder, which gives an estimate of the transmitted information sequence $\un{\hat{\bU}}$.

We consider the discrete-time, memoryless, vectorial AWGN channel 
\begin{align}\label{awgn}
\bY_{n}=\bX_{n}+\bZ_{n}
\end{align} 
where $\bX_{n}, \bY_{n}, \bZ_{n}$ are $4$D real vectors and $n=1,2,\ld,\Ns$ is the discrete-time index. The components of the noise vector $\bZ_{n}$ are independent, zero-mean, Gaussian random variables with variance $\No/2$ in each dimension, and thus,
\begin{align}\label{awgn.pdf}
f_{\bY_{n}|\bX_{n}}(\by|\bx) = \frac{1}{(\pi \No)^{2}}\exp{\left(-\frac{\|\by-\bx\|^{2}}{\No}\right)}.
\end{align} 

The communication channel in \figref{model} encompasses all the transmitter digital signal processing (DSP) used after the bit-to-symbol mapping (i.e., pulse shaping, polarization multiplexing, filtering, electro-optical conversion, etc.), the physical channel (the fiber, amplifiers, regenerators, etc.), and the receiver DSP (optical-to-electrical conversion, filtering, equalization, digital back-propagation, matched filtering, etc.). The use of the AWGN channel in \eqref{awgn} to model all these blocks can be justified in amplified spontaneous emission noise dominated links where chromatic dispersion and polarization mode dispersion are perfectly compensated. The AWGN assumption also holds for uncompensated coherent systems where the so-called \emph{GN model} has been widely used (see \cite{Poggiolini14a} and references therein).

At each time instant $n$, the transmitted vector $\bX_{n}$ is selected with equal probability from a constellation $\mcS\triangleq\set{\bs_1,\bs_2,\ld,\bs_{M}}$, where $M=2^m$. The average symbol energy is $\Es\triangleq \E[\|\bX\|^2]=(1/M)\sum_{i=1}^{M}\|\bs_{i}\|^2$ and the SNR is defined as $\gamma\triangleq \Es/\No$. For a rate $\Rc$ FEC encoder, the spectral efficiency in bits/symbol is $\R=\Rc m$. The length of the information sequence $\un{\bU}$ is $\Nb=\eta\Ns$ and the average bit energy is $\Eb=\Es/\R$.

The transmitter in \figref{model} is a one-to-one mapping between the information sequence $\un{\bU} \in \set{0,1}^{\Nb}$ and the coded sequence $\un{\bX} \in \mcC \subseteq \mcS^{\Ns}$, where $|\mcC| = 2^{\Nb}$. The set $\mcC$ is called the \emph{codebook}, and the mapping between the $2^{\Nb}$ information sequences and the code $\mcC$ is called the CM \emph{encoder}. At the receiver side, a CM \emph{decoder} (see \figref{model}) uses the mapping rule used at the transmitter (as well as the channel characteristics) to give an estimate of the information sequence. The triplet codebook, encoder, and decoder forms a so-called \emph{coding scheme}. Practical coding schemes are designed so as to minimize the probability that $\hat{\un{\bU}}$ differs from $\un{\bU}$, while at the same time keeping the complexity of both encoder and decoder low.

\subsection{CM Decoder Structures}\label{Sec:Model:Rx}

\begin{figure}[t]
\begin{center}
\newcommand{\scale}{0.82}
\psfrag{bi}[br][Br][\scale]{$\un{\bU}$}
\psfrag{ENCODER}[Bc][Bc][\scale]{CM Encoder}
\psfrag{x}[bc][Bc][\scale]{$\un{\bX}$}
\psfrag{z}[bc][Bc][\scale]{$\un{\bZ}$}
\psfrag{ddd}[cc][cc][\scale]{$\vdots$}
\psfrag{y}[bc][Bc][\scale]{$\un{\bY}$}
\psfrag{ML}[Bc][Bc][\scale]{Optimum Decoder}
\psfrag{DECML1}[cc][cc][\scale]{ML Decoder}
\psfrag{DECML2}[cc][cc][\scale]{$\argmax_{\un{\bx}} f_{\un{\bY}|\un{\bX}}(\un{\by}|\un{\bx})$}
\psfrag{BI}[Bc][Bc][\scale]{BW Decoder}
\psfrag{AWGN}[Bc][Bc][\scale]{AWGN Channel}
\psfrag{LLR1}[cc][cc][\scale]{}
\psfrag{LLR2}[ll][cc][\scale]{\overlay{-7.2mm}{1mm}{Demapper}}
\psfrag{lc1}[cc][cc][\scale]{$\un{\Lambda}_{1}$}
\psfrag{lcm}[cc][cc][\scale]{$\un{\Lambda}_{m}$}
\psfrag{DEC1}[cl][cc][\scale]{\overlay{-4.3mm}{1mm}{Binary}}
\psfrag{DEC2}[lc][cc][\scale]{\overlay{-5.5mm}{1mm}{SD-FEC}\overlay{-5.5mm}{-3mm}{Decoder}}
\psfrag{bih}[bl][Bl][\scale]{$\un{\hat{\bU}}$}
\includegraphics[width=0.85\columnwidth]{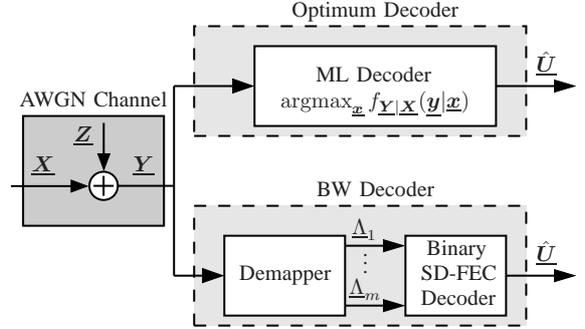}
\end{center}
\caption{Two implementations of the CM decoder in \figref{model}: Optimum (ML) decoder (top) and BW decoder (bottom).}
\label{receivers}
\end{figure}

\figref{receivers} shows two possible receiver structures for the CM encoder in \figref{model} together with the AWGN channel in \eqref{awgn.pdf}: the optimal ML decoder and the (suboptimal) BW decoder. The ML decoder operates on the sequence of symbols $\un{\bY}$ and finds the most likely coded sequence, i.e., it performs $\un{\hat{\bu}}=\argmax_{\un{\bx}} f_{\un{\bY}|\un{\bX}}(\un{\by}|\un{\bx})$. On the other hand, the BW decoder computes soft information on the code bits $\un{B}_1, \ldots, \un{B}_m$ on a symbol-by-symbol basis. This soft information is typically represented in the form of LLRs $\un{\Lambda}_1, \ldots, \un{\Lambda}_m$, where $\un{\Lambda}_k = [\Lambda_{1,k},\ldots,\Lambda_{\Ns,k}]$ for $k=1,\ldots,m$. These LLRs are then passed to a binary SD-FEC decoder.\footnote{Alternatively, an HD demapper can be combined with an HD-FEC decoder. In this paper, we only consider SD-FEC decoders.}

Assuming perfect knowledge of $\No$, at each discrete-time instant $n$, $m$ LLRs are calculated as
\begin{align}\label{LLR.def}
\Lambda_{n,k} 	& \triangleq \log \frac{f_{\bY_n|B_{n,k}}(\by|1)}{f_{\bY_n|B_{n,k}}(\by|0)}\\
\label{LLR.1}
			& = \log\frac{\sum_{\bs\in\mcSko}\exp(-\frac{1}{\No}\|\by-\bs\|^{2})}{\sum_{\bs\in\mcSkz}\exp(-\frac{1}{\No}\|\by-\bs\|^{2})}\\
\label{LLR.max-log}
			& \approx \frac{1}{\No} \left(\min_{\bs\in\mcSkz}\|\by-\bs\|^{2}-\min_{\bs\in\mcSko}\|\by-\bs\|^{2}\right)
\end{align} 
where \eqref{LLR.1} follows from \eqref{awgn.pdf} and $\mcSkb\subset\mcS$ is the set of constellation symbols labeled with a bit $b\in\set{0,1}$ at bit position $k\in\set{1,\ld,m}$. The approximation in \eqref{LLR.max-log} follows from using the so-called max-log approximation\cite{Viterbi98}. 

Alternatively, the LLRs in \eqref{LLR.def} can be defined as 
\begin{align}\label{LLR.def.alt}
\Lambda_{n,k} 	& = \log \frac{f_{\bY_n|B_{n,k}}(\by|0)}{f_{\bY_n|B_{n,k}}(\by|1)},
\end{align}
which could have some advantages in practical implementations. For example, in a popular complement-to-two binary format, the most significant bit carries the sign, i.e., when an MSB is equal to zero (0), it means that a number is positive, and when an MSB is equal to one (1), it means that the number is negative. Then, if \eqref{LLR.def.alt} is used, the transmitted bit obtained via HDs can be recovered directly from the MSB. 

Without loss of generality, in this paper we use the definition in \eqref{LLR.def}. Furthermore, since the mapper, channel, and demapper are all memoryless, the time index $n$ is dropped from now on.

Throughout this paper we denote the pre-FEC BER and the post-FEC BER by $\BERin$ and $\BERout$, respectively. $\BERin$ can be obtained from the max-log LLRs in \eqref{LLR.max-log} as \cite[Theorem~1]{Ivanov12a}
\begin{align}\label{BERin.def}
\BERin = \frac{1}{m}\sum_{k=1}^{m} \frac{1}{2}\sum_{b\in\set{0,1}}\int_{0}^{\infty}
f_{\Lambda_{k}|B_{k}}((-1)^{b}\lambda |b)\,\tnr{d}\lambda
\end{align}
and depends only on the constellation, its binary labeling and the communication channel.\footnote{Note that HDs on the exact LLRs in \eqref{LLR.1} give slightly worse pre-FEC BER results in the low-SNR regime.} On the other hand, $\BERout$ also depends on the choice of FEC code.

The BW decoder in \figref{receivers} is usually known as a BICM receiver/decoder, owing its name to the original works \cite{Zehavi92,Caire98}, where a bit-level interleaver was included between the FEC encoder and mapper. We refrain from using such a name because the interleaver might or might not be included, and if included, it can be assumed to be part of the FEC encoder.

\subsection{Achievable Rates}\label{Sec:Model:AR}

A rate $R$ (in bits/symbol) is said to be \emph{achievable} at block length $\Ns$ and average error probability $\epsilon$ if there exists a coding scheme, consisting of a codebook $\mcC$, an encoder, and a decoder, such that $|\mcC| = 2^{R \Ns}$ and $\Pr\{\un{\hat{\bU}} \ne \un{\bU}\} \le \epsilon$. The largest achievable rate at given $\Ns$ and $\epsilon$ is denoted by $R^{*}(\Ns,\epsilon)$. The \emph{channel capacity} $C$ is the largest achievable rate for which a coding scheme with vanishing error probability exists, in the limit of large block length \cite[Sec. 1 and 14]{Shannon48}, i.e.,
\begin{align}\label{C.def}
C \triangleq \lim_{\epsilon\to 0}\lim_{\Ns\to\infty}R^{*}(\Ns,\epsilon).
\end{align}
The channel capacity is often defined subject to an average power constraint $P$, which means that every codeword $\un{\bX}=[\bX_{1},\ld,\bX_{\Ns}] \in \mcC$ must satisfy $\sum_n \|\bX_{n}\|^2 \le P$.

For memoryless channels and a given constellation $\mcS$, the largest achievable rate is the MI between $\bX$ and $\bY$ defined as
\begin{align}\label{MI}
I(\bX;\bY) & \triangleq \E\left[\log_{2}\frac{f_{\bY|\bX}(\bY|\bX)}{f_{\bY}(\bY)}\right].
\end{align}
By Shannon's \emph{channel coding theorem,} the channel capacity of a discrete-time memoryless channel with an average power constraint can be calculated as \cite{Shannon48}, \cite[Ch.~7]{Cover06_Book}
\begin{align}\label{C.MI}
C = \sup_{f_{\bX}:\Es \leq P} I(\bX;\bY)
\end{align}
where $I(\bX;\bY)$ is the MI in \eqref{MI}  and the maximization in \eqref{C.MI} is over all distributions\footnote{In general, the capacity-achieving distribution can be discrete, continuous, or mixed.} of $\bX$ that satisfy the average power constraint $\Es \leq P$, for a given channel $f_{\bY|\bX}$. For the $4$D channel in \eqref{awgn}, \eqref{C.MI} gives
\begin{align}\label{C.AWGN}
C = \frac{N}{2}\log_{2}{\left(1+\frac{2}{N}\gamma\right)}=2\log_{2}{\left(1+\frac{\gamma}{2}\right)}
\end{align}
which is attained by a zero-mean Gaussian input distribution $f_{\bX}$ with a diagonal covariance matrix with all diagonal entries equal to $\Es/4=P/4$.

In this paper, we consider equally likely symbols and discrete constellations $\mcS$, and thus, $f_{\bX}$ is a uniform distribution over $\mcS$. In this case, the MI in \eqref{MI} becomes
\begin{align}\label{MI.uniform}
I(\bX;\bY) & = \frac{1}{M} \sum_{\bs\in\mcS}  \int_{\Real^{4}}f_{\bY|\bX}(\by|\bs) \log_{2}\frac{f_{\bY|\bX}(\by|\bs)}{f_{\bY}(\by)}\, \tnr{d}\by.
\end{align}
The MI $I(\bX;\bY)$ in \eqref{MI.uniform} is the largest achievable rate for the optimum ML decoder and a given constellation $\mcS$. Thus, for the optimal ML decoder, reliable transmission with arbitrarily low error probability is possible if $\R< I(\bX;\bY)$. By Shannon's random coding paradigm, the rate in \eqref{MI.uniform} is achievable using a codebook $\mcC$ consisting of $2^{R \Ns}$ codewords of length $\Ns$, each symbol drawn independently and uniformly from $\mcS$.

When the BW decoder in \figref{receivers} is considered, due to the fact that this decoder is not ML, the largest achievable rate is unknown. The most popular achievable rate for the BW decoder is the GMI\footnote{The GMI is not necessarily the \emph{largest} achievable rate for the receiver in \figref{receivers}. Other achievable rates include the so-called LM rate \cite[Part~I]{Peng12_Thesis} and the newly derived rate for nonequally likely constellation points (i.e., probabilistic shaping) \cite[Theorem~1]{Bocherer14}.} \cite{Martinez09}
\begin{align}\label{GMI}
I^{\tnr{gmi}} = \sum_{k=1}^{m}I(B_{k};\bY)
\end{align}
where
\begin{align}\label{BW.MI}
I(B_{k};\bY)= \E\left[\log_2{\frac{f_{\bY|B_{k}}(\bY|B_{k})}{f_{\bY}(\bY)}}\right].
\end{align}

In analogy with \eqref{MI.uniform}, we consider in this paper independent, equally likely bits, in which case the GMI in \eqref{GMI} becomes
\begin{align}
I^{\tnr{gmi}} = \frac{1}{2} \sum_{k=1}^m\sum_{b\in\{0,1\}}  \int_{\Real^{4}}f_{\bY|B_k}(\by|b) \log_{2}\frac{f_{\bY|B_k}(\by|b)}{f_{\bY}(\by)}\, \tnr{d}\by.
\end{align}
This rate is achievable with a BW decoder, without iterative decoding, using the same codebook $\mcC$ that achieves \eqref{MI.uniform} with an optimum decoder. Note that designing a codebook by drawing symbols independently and uniformly from $\mcS$ corresponds to independent and equally likely bits $B_k$.

When the LLRs are calculated using \eqref{LLR.1}, it can be shown that \cite[Theorem~4.21]{Alvarado15_Book}
\begin{align}\label{GMI.LLR}
I^{\tnr{gmi}} = \sum_{k=1}^{m}I(B_{k};\Lambda_{k}) = \sum_{k=1}^{m}\E\left[\log_2{\frac{f_{\Lambda_{k}|B_{k}}(\Lambda_{k}|B_{k})}{f_{\Lambda_{k}}(\Lambda_{k})}}\right].
\end{align}
When the LLRs are calculated using \eqref{LLR.max-log}, the resulting achievable rate is smaller than $I^{\tnr{gmi}}$ in \eqref{GMI.LLR}. Under certain conditions, this loss can be recovered by correcting the max-log LLRs, as shown in \cite{Jalden10,Nguyen11} (see also \cite[Ch.~7]{Alvarado15_Book}).

Achievable rates for BW decoders were first analyzed in \cite{Caire98}. The BW decoder was later recognized in \cite{Martinez09} as a mismatched decoder, where it was shown that the GMI in \eqref{GMI} is an achievable rate. It was also shown in \cite{Martinez09} that in terms of achievable rates, the interleaver plays no role, and that the key element is the suboptimal (mismatched) decoder.

The GMI in \eqref{GMI} is an achievable rate for BW decoders but has not been proven to be the largest achievable rate. Finding the largest achievable rate remains as an open research problem. Despite this cautionary statement, the GMI has been shown to predict very well the performance of BW decoders based on capacity-approaching FEC encoder--decoder pairs. This has been shown for example in \cite[Sec.~V]{Alvarado06c}, \cite[Sec.~V-D]{Fertl12}, and \cite[Sec.~IV]{Alvarado14a}. Generally speaking, when good turbo or LDPC codes are used, the gap between the coded system and the GMI prediction is usually less than $1$~dB.

The mapper is one-to-one, and thus, $I(\bB;\bY) = I(\bX;\bY)$. The chain rule of MI \cite[Sec.~2.5]{Cover06_Book} gives 
\begin{align}\label{Chain.Rule}
I(\bB;\bY)\geq\sum_{k=1}^{m}I(B_{k};\bY)
\end{align}
and thus,
\begin{align}\label{MI.vs.GMI}
I^{\tnr{gmi}} \leq I(\bX;\bY).
\end{align}
The difference $I(\bX;\bY) - I^{\tnr{gmi}}$ can be understood as the loss in terms of achievable rates caused by the use of a BW decoder. Furthermore, the GMI (unlike the MI) is highly dependent on the binary labeling. Gray codes are known to be good for high SNR \cite[Fig.~4]{Caire98}, \cite{Agrell10b}, \cite[Sec.~IV]{Alvarado11b}, but for many constellations, they do not exist.

Closed-form expressions for the MI and GMI are in general unknown, and thus, numerical methods are needed. For the AWGN channel, both MI and GMI can be efficiently calculated based on Gauss--Hermite quadrature. To this end, the ready-to-use expressions in \cite[Sec.~III]{Alvarado11b} can be used. The GMI can also be calculated using the approximation recently introduced in \cite{Alvarado14a}. This approximation is particularly useful to find good binary labelings in terms of GMI. When the channel is unknown or when the dimensionality of the constellation grows, Monte Carlo integration is preferred.

\section{Post-FEC BER Prediction via GMI}\label{Sec:NumResults:GMI}

In this section, we consider the problem of predicting the decoder's performance for a given code rate. To this end, we first introduce the concept of the \emph{BICM channel} (see \cite[Fig.~1]{Martinez06}, \cite[Fig.~1]{Alvarado07d}). The BICM channel\footnote{Also called ``modulation channel'' in \cite[Fig.~1]{Fertl12}.} encompasses all the elements that separate the encoder and decoder (see Figs.~\ref{model}~and~\ref{receivers}), i.e., the mapper and demapper, transmitter and receiver DSP, fiber, amplifiers, filtering, equalization, etc. The BICM channel is then what the encoder--decoder pair ``sees''.

In principle, to predict the post-FEC BER of a given encoder over different BICM channels (e.g., different constellations, different amplification schemes, different fiber types, etc.), the whole communication chain should be re-simulated. To avoid this, one could try to find an easy-to-measure metric that characterizes the BICM channel and hope that different channels with the same metric result in the same $\BERout$. Here we consider four different metrics and argue that the GMI in \eqref{GMI} is the most appropriate one.

Consider the irregular repeat-accumulate LDPC codes proposed by the second generation satellite digital video broadcasting standard \cite{ETSI_EN_302_307_v121} and the 6 code rates
\begin{align}\label{coding.rates}
\Rc \in \set{1/3, 2/5, 1/2, 3/5, 3/4, 9/10}
\end{align}
which correspond to the FEC overheads $\set{200,150,100,66.6,33.3,11.1}$\%. Each transmitted block consists of $64\,800$ code bits which are randomly permuted before being cyclically assigned to the binary sequences $\un{B}_1, \ld, \un{B}_m$. At the receiver, LLRs $\Lambda_{k}$ are calculated using \eqref{LLR.1} and passed to the SD-FEC decoder, which performs 50 iterations. 

\begin{figure}[tbp]
\newcommand{\scale}{0.8}
\newcommand{\scalesmall}{0.65}
\centering
\psfrag{PM-QPSK}[cc][cc][\scale]{{\fcolorbox{white}{white}{{PM-QPSK}}}}%
\psfrag{PM-16QAM}[cc][cc][\scale]{{\fcolorbox{white}{white}{{PM-16QAM}}}}%
\psfrag{PM-64QAM}[cc][cc][\scale]{{\fcolorbox{white}{white}{{PM-64QAM}}}}%
\psfrag{PM-256QAM}[cc][cc][\scale]{{\fcolorbox{white}{white}{{PM-256QAM}}}}%
\psfrag{r13}[cc][cc][\scale][-85]{{\fcolorbox{white}{white}{{$\Rc=1/3$}}}}%
\psfrag{r25}[cc][cc][\scale][-85]{{\fcolorbox{white}{white}{{$\Rc=2/5$}}}}%
\psfrag{r12}[cc][cc][\scale][-85]{{\fcolorbox{white}{white}{{$\Rc=1/2$}}}}%
\psfrag{r35}[cc][cc][\scale][-85]{{\fcolorbox{white}{white}{{$\Rc=3/5$}}}}%
\psfrag{r34}[cc][cc][\scale][-85]{{\fcolorbox{white}{white}{{$\Rc=3/4$}}}}%
\psfrag{r91}[cc][cc][\scale][-85]{{\fcolorbox{white}{white}{{$\Rc=9/10$}}}}%
\psfrag{ylabel}[cc][cb][\scale]{$\BERout$}%
\psfrag{xlabel}[cc][tc][\scale]{SNR $\gamma$~[dB]}%
\includegraphics{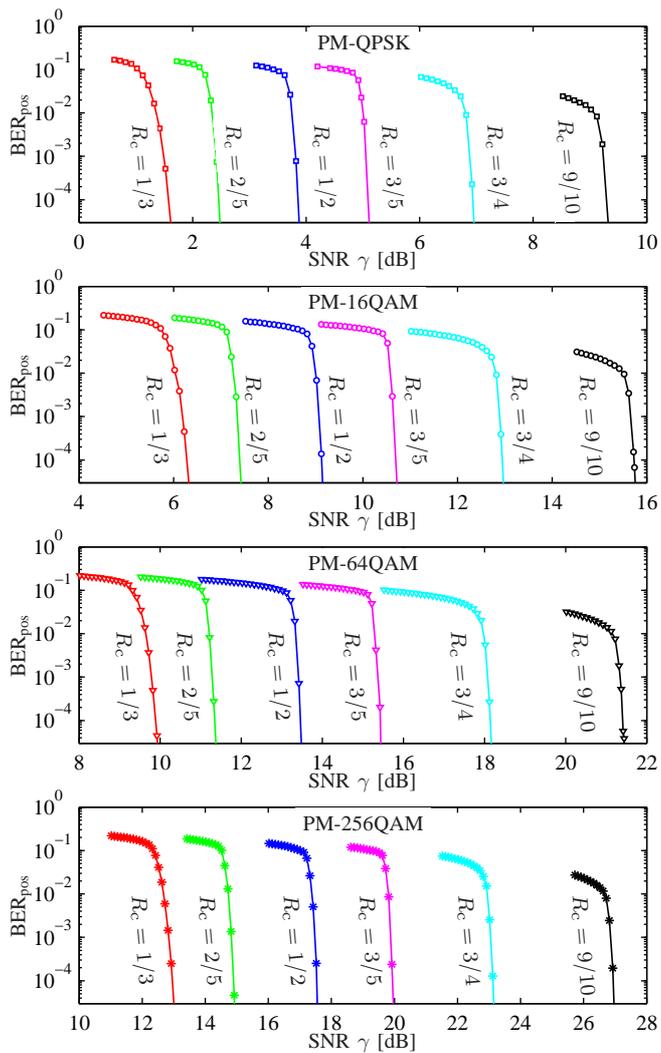}%
\overlay{-6.9cm}{9.2cm}{\rotatebox{-85}{\footnotesize$\Rc=1/3$}}%
\overlay{-5.6cm}{9.2cm}{\rotatebox{-85}{\footnotesize$\Rc=2/5$}}%
\overlay{-4.5cm}{9.2cm}{\rotatebox{-85}{\footnotesize$\Rc=1/2$}}%
\overlay{-3.5cm}{9.2cm}{\rotatebox{-85}{\footnotesize$\Rc=3/5$}}%
\overlay{-2.1cm}{9.2cm}{\rotatebox{-85}{\footnotesize$\Rc=3/4$}}%
\overlay{-1.2cm}{9.2cm}{\rotatebox{-85}{\footnotesize$\Rc=9/10$}}%
\overlay{-7.3cm}{5.8cm}{\rotatebox{-85}{\footnotesize$\Rc=1/3$}}%
\overlay{-6.5cm}{5.8cm}{\rotatebox{-85}{\footnotesize$\Rc=2/5$}}%
\overlay{-5.3cm}{5.8cm}{\rotatebox{-85}{\footnotesize$\Rc=1/2$}}%
\overlay{-4.3cm}{5.8cm}{\rotatebox{-85}{\footnotesize$\Rc=3/5$}}%
\overlay{-2.9cm}{5.8cm}{\rotatebox{-85}{\footnotesize$\Rc=3/4$}}%
\overlay{-1.3cm}{5.8cm}{\rotatebox{-85}{\footnotesize$\Rc=9/10$}}%
\overlay{-7.1cm}{2.3cm}{\rotatebox{-85}{\footnotesize$\Rc=1/3$}}%
\overlay{-6.2cm}{2.3cm}{\rotatebox{-85}{\footnotesize$\Rc=2/5$}}%
\overlay{-5.1cm}{2.3cm}{\rotatebox{-85}{\footnotesize$\Rc=1/2$}}%
\overlay{-4.1cm}{2.3cm}{\rotatebox{-85}{\footnotesize$\Rc=3/5$}}%
\overlay{-2.8cm}{2.3cm}{\rotatebox{-85}{\footnotesize$\Rc=3/4$}}%
\overlay{-1.3cm}{2.3cm}{\rotatebox{-85}{\footnotesize$\Rc=9/10$}}%
\caption{Post-FEC BER ($\BERout$) for different code rates $\Rc$ and constellations as a function of the SNR $\gamma$. The constellations are PM-QPSK (squares), PM-16QAM (circles), PM-64QAM (triangles), and PM-256QAM (stars).}
\label{BERout_vs_SNR}
\end{figure}

\figref{BERout_vs_SNR} shows the performance of the LDPC decoder with PM-QPSK, PM-16QAM, PM-64QAM, and PM-256QAM as a function of SNR. There are $24$ different coding and modulation pairs, leading to $24$ spectral efficiencies $\R=\Rc m$. The results in this figure show that, for any given code rate, different modulations have very different SNR requirements. For example, for $\Rc=3/5$ and a target post-FEC BER of $10^{-4}$, the SNR thresholds are $5.1$~dB, $10.8$~dB, $15.5$~dB and $20$~dB for PM-QPSK, PM-16QAM, PM-64QAM, and PM-256QAM, respectively. This leads to the obvious conclusion that SNR cannot be used to predict the post-FEC BER performance of a given code when used with different constellations.

Under some assumptions on independent errors within a block,\footnote{This can be guaranteed by properly interleaving the code bits.} the pre-FEC BER in \eqref{BERin.def} can be used to predict the post-FEC BER of HD-FEC decoders.
Based on such relations, the conventional design paradigm in optical communications is to design systems for a certain required pre-FEC BER, the so-called \emph{FEC limit} or FEC threshold, which is typically in the range $10^{-4} - 10^{-3}$. The HD-FEC decoder is then \emph{assumed} to bring down the post-FEC BER to, say, $10^{-12}$ or $10^{-15}$, without actually including any coding in simulations or experiments.

For a given (fixed) BICM channel, the pre-FEC BER can also be used to predict the post-FEC BER of an SD-FEC decoder. This has been done for example for some of the SD-FEC decoders in the G.975.1 standard \cite{ITU-T_G.975.1}, where post-FEC BER values are given as a function of pre-FEC BER. There is nothing fundamentally wrong with presenting post-FEC BER as a function of pre-FEC BER. However, more often than not, reported uncoded experiments or simulations rely on these tabulated values and claim (without encoding and decoding information) the existence of an SD-FEC decoder that can deal with the measured pre-FEC BER. The caveat with this approach is that it relies on the strong assumption that the same SD-FEC encoder and decoder pair will perform identically for two different BICM channels which happen to have the same pre-FEC BER.

\begin{figure}[tbp]
\newcommand{\scale}{0.8}
\newcommand{\scalesmall}{0.65}
\centering
\psfrag{ylabel}[cc][cb][\scale]{$\BERout$}%
\psfrag{13}[cc][cc][\scale][-90]{{\fcolorbox{white}{white}{{$\Rc=1/3$}}}}%
\psfrag{25}[cc][cc][\scale][-90]{{\fcolorbox{white}{white}{{$\Rc=2/5$}}}}%
\psfrag{12}[cc][cc][\scale][-90]{{\fcolorbox{white}{white}{{$\Rc=1/2$}}}}%
\psfrag{35}[cc][cc][\scale][-90]{{\fcolorbox{white}{white}{{$\Rc=3/5$}}}}%
\psfrag{34}[cc][cc][\scale][-90]{{\fcolorbox{white}{white}{{$\Rc=3/4$}}}}%
\psfrag{91}[cc][cc][\scale][-90]{{\fcolorbox{white}{white}{{$\Rc=9/10$}}}}%
\psfrag{xlabel}[cc][cB][\scale]{$1-\BERin$}%
\includegraphics{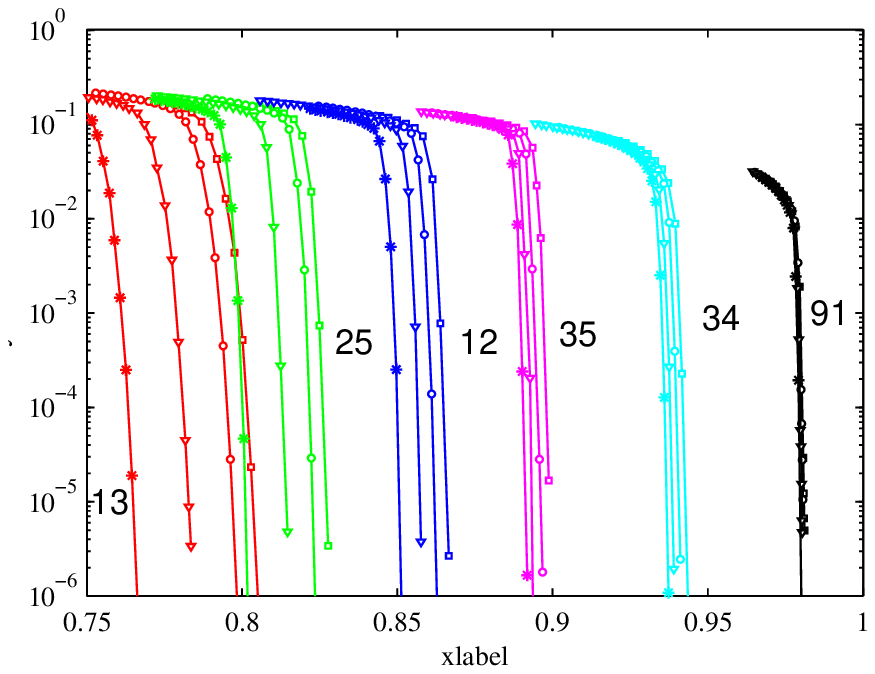}
\caption{Post-FEC BER ($\BERout$) as a function of pre-FEC BER ($\BERin$) for the $24$ cases in \figref{BERout_vs_SNR}. The same markers are used.}
\label{BERout_vs_BERin}
\end{figure}

To study the robustness of the pre-FEC BER as a metric to predict post-FEC BER, we show in \figref{BERout_vs_BERin} $\BERout$ as a function of $\BERin$ for the same $24$ combinations of constellations and codes as in \figref{BERout_vs_SNR}. Ideally, all lines corresponding to the same code rate should fall on top of each other, indicating that measuring $\BERin$ is sufficient to predict the post-FEC BER when the BICM channel changes (in this case, due to the change in modulation format). The results in this figure show that the curves get ``grouped'' for the same code rate, and thus, $\BERin$ is a better metric than SNR (cf.~\figref{BERout_vs_SNR}). The results in \figref{BERout_vs_BERin} also show that $\BERin$ is a good metric for very high code rates. For low and moderate code rates, however, $\BERin$ fails to predict the performance of the decoder. The implication of this is that measuring pre-FEC BER \emph{cannot} be used to predict the post-FEC BER of an encoder--decoder pair across different BICM channels. The FEC-limit design paradigm fails.

\begin{figure}[tbp]
\newcommand{\scale}{0.8}
\newcommand{\scalesmall}{0.65}
\centering
\psfrag{ylabel}[cc][cb][\scale]{$\BERout$}%
\psfrag{xlabel}[cc][cB][\scale]{$I(\bX;\bY)/m$}%
\includegraphics{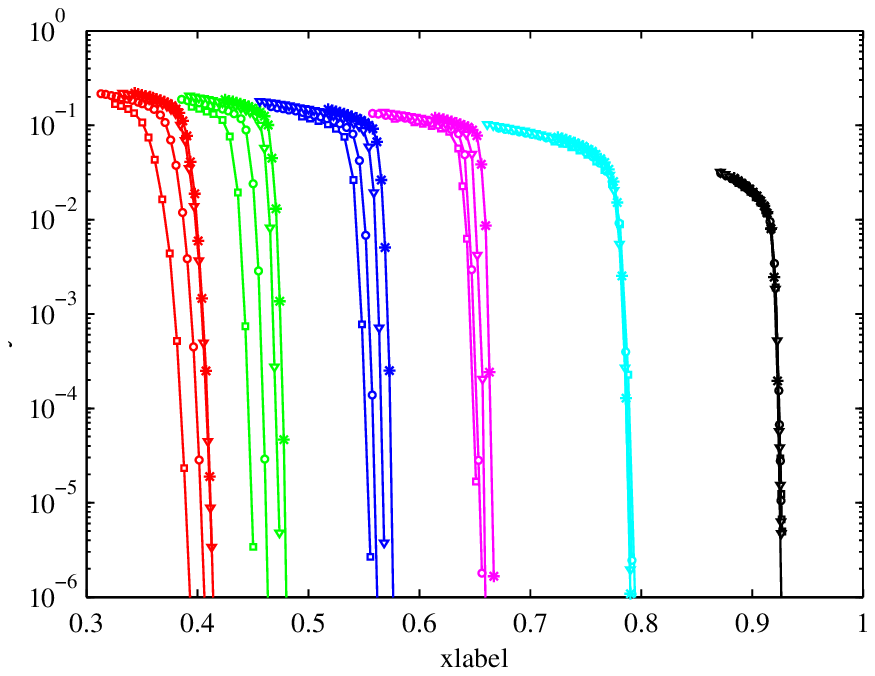}%
\overlay{-7.8cm}{4cm}{\rotatebox{-90}{\footnotesize$\Rc=1/3$}}%
\overlay{-6.1cm}{4cm}{\rotatebox{-90}{\footnotesize$\Rc=2/5$}}%
\overlay{-5.0cm}{4cm}{\rotatebox{-90}{\footnotesize$\Rc=1/2$}}%
\overlay{-4.0cm}{4cm}{\rotatebox{-90}{\footnotesize$\Rc=3/5$}}%
\overlay{-2.6cm}{4cm}{\rotatebox{-90}{\footnotesize$\Rc=3/4$}}%
\overlay{-1.0cm}{4cm}{\rotatebox{-90}{\footnotesize$\Rc=9/10$}}%
\caption{Post-FEC BER ($\BERout$) as a function of the normalized MI ($I(\bX;\bY)/m$) for the $24$ cases in \figref{BERout_vs_SNR} (same markers).}
\label{BERout_vs_MI}
\end{figure}

In \figref{BERout_vs_MI}, we consider $\BERout$ as a function of the (normalized) MI. The obtained results indicate that the MI is slightly better than $\BERin$ at predicting $\BERout$ (the curves for low code rates are more compact). The same trend was observed in \cite{Leven11} (for a BW decoder with differentially encoded PM-QPSK), where the idea of using MI instead of $\BERin$ was first introduced. As explained in \secref{Sec:Model:AR}, however, the MI is in principle not connected to the performance of a BW decoder, which may explain why the curves are still significantly spread out, particularly at lower code rates.

\begin{figure}[tbp]
\newcommand{\scale}{0.8}
\newcommand{\scalesmall}{0.65}
\centering
\psfrag{ylabel}[cc][cb][\scale]{$\BERout$}%
\psfrag{xlabel}[cc][cB][\scale]{$I^{\tnr{gmi}}/m$}%
\includegraphics{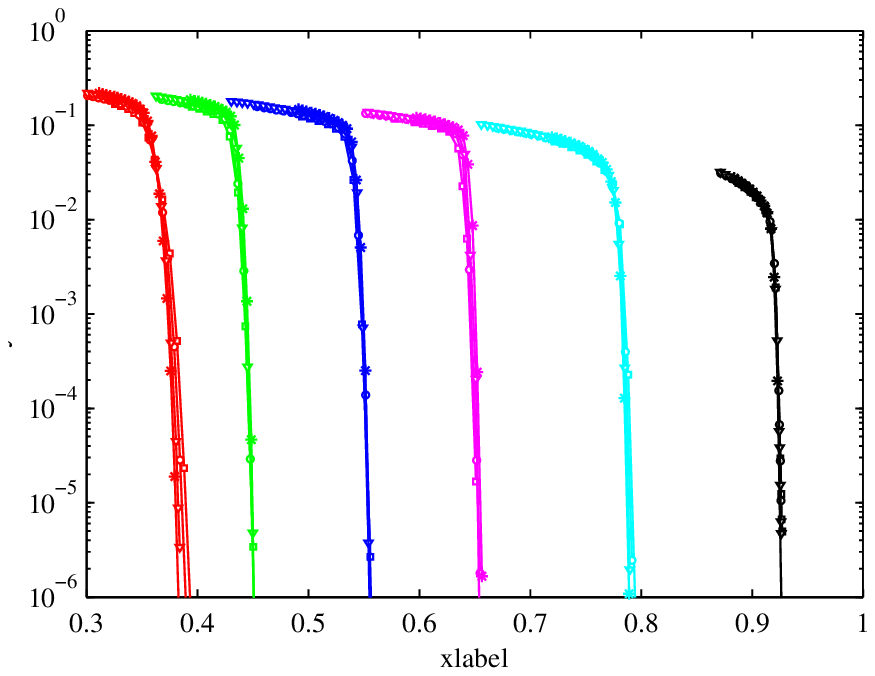}%
\overlay{-7.9cm}{4cm}{\rotatebox{-90}{\footnotesize$\Rc=1/3$}}%
\overlay{-6.4cm}{4cm}{\rotatebox{-90}{\footnotesize$\Rc=2/5$}}%
\overlay{-5.3cm}{4cm}{\rotatebox{-90}{\footnotesize$\Rc=1/2$}}%
\overlay{-4.1cm}{4cm}{\rotatebox{-90}{\footnotesize$\Rc=3/5$}}%
\overlay{-2.6cm}{4cm}{\rotatebox{-90}{\footnotesize$\Rc=3/4$}}%
\overlay{-1.0cm}{4cm}{\rotatebox{-90}{\footnotesize$\Rc=9/10$}}%
\caption{Post-FEC BER ($\BERout$) as a function of the normalized GMI ($I^{\tnr{gmi}}/m$).}
\label{BERout_vs_GMI}
\end{figure}

Based on the analysis in \secref{Sec:Model:AR}, we propose here to study $\BERout$ as a function of the GMI. The information-theoretic rationale behind this idea is that a SD-FEC decoder is fed with LLRs, and thus, the GMI is a better metric (see \eqref{GMI.LLR}). The values of $\BERout$ as a function of the GMI are shown in \figref{BERout_vs_GMI}.\footnote{The MIs and GMIs were estimated using Monte Carlo integration.} These results show that for any given code rate, changing the constellation does not greatly affect the post-FEC BER prediction if the GMI is kept constant. More importantly, and unlike for the pre-FEC BER, the prediction based on the GMI appears to work across all code rates. 

The results in \figref{BERout_vs_GMI} suggest that measuring the GMI of the BICM channel is the correct quantity to characterize the post-FEC BER of a (capacity-achieving) SD-FEC decoder. Although for high code rates the results in \figref{BERout_vs_GMI} are somehow similar to those in \figsref{BERout_vs_BERin}{BERout_vs_MI}, we have no theoretical justification the use $\BERin$ or MI as a metric to predict the performance of a SD-FEC. More importantly, having a metric like the GMI that works for all code rates is very important. Considering only high code rates---as is usually done in the optical community---is an artificial constraint that reduces flexibility in the design, as correctly pointed out in \cite[Sec.~II-B]{Smith10}.

\section{Achievable Rates}\label{Sec:NumResults}

In this section, we focus on cases where the number of bits per dimension is an integer, due to their practical relevance. The examples studied have $1$, $2$, and $3$ bits/dimension, which corresponds to, respectively, $4$, $8$, and $12$ bits/symbol or $M=16$, $256$, and $4096$ constellation points.

\subsection{Achievable Rates for $M=16$}\label{M.16}

We consider three $4$D constellations with $M=16$: PM-QPSK, $\Cfs$, and SO-PM-QPSK. While $\Cfs$ is asymptotically the best constellation in terms of $\BERin$, PM-QPSK and SO-PM-QPSK have the advantage of a lower implementation complexity. On the other hand, the results in \cite{Karout2013_OFC, Bulow2013} show that $\Cfs$ gives higher MI than PM-QPSK at all SNRs. This indicates that $\Cfs$ is the best choice among these constellations for capacity-approaching CM transmitters with ML decoding. 

In terms of binary labelings, we use the unique Gray code for PM-QPSK, which assigns a separate bit to each dimension. Thus, PM-QPSK becomes the Cartesian product of four binary shift keying (BPSK) constellations, $\sum_{k=1}^{m}I(B_{k};\bY) = I(\bX;\bY)$, and thus, \eqref{MI.vs.GMI} holds with equality. In other words, PM-QPSK causes no penalty in terms of achievable rates if a BW decoder is used. For SO-PM-QPSK, we use the labeling proposed in \cite{Sjodin2013_CL}, while for $\Cfs$ we use a labeling (found numerically) that gives high GMI for a wide range of SNR.

\begin{figure}[tbp]
\newcommand{\scale}{0.8}
\newcommand{\scalesmall}{0.65}
\centering
\psfrag{xlabel}[cc][cB][\scale]{$\Eb/\No$~[dB]}%
\psfrag{ylabel}[cc][cb][\scale]{$\R$~[bits/symbol]}%
\psfrag{CapAWGNCapacity}[cl][cl][\scalesmall]{Channel Capacity}%
\psfrag{PMQMI}[cl][cl][\scalesmall]{PM-QPSK (G)MI}%
\psfrag{MI}[cl][cl][\scalesmall]{MI}%
\psfrag{GMI}[cl][cl][\scalesmall]{GMI}%
\psfrag{C416}[cl][cl][\scale]{$\Cfs$}%
\psfrag{SO-PM-QPSK}[cl][cl][\scale]{SO-PM-QPSK}%
\psfrag{PM-QPSK}[cl][cl][\scale]{PM-QPSK}%
\includegraphics[width=\columnwidth]{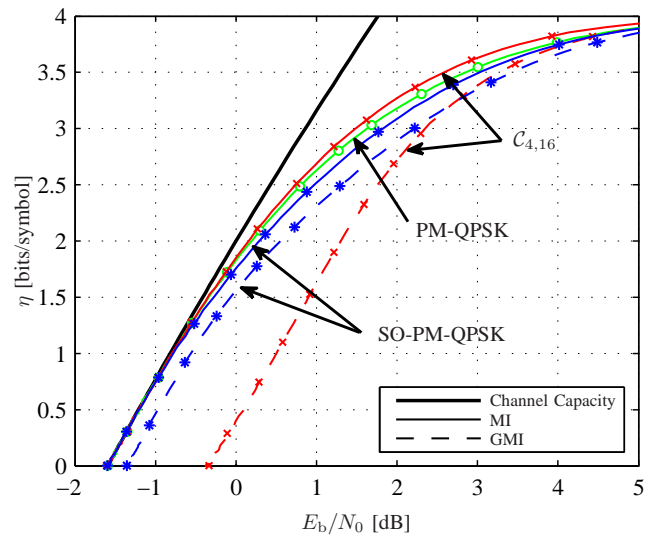}
\caption{MI and GMI for three constellations with $M=16$: PM-QPSK (circles), $\Cfs$ (crosses), and SO-PM-QPSK (stars). The MI and GMI overlap for PM-QPSK. The channel capacity in \eqref{C.AWGN} is also shown (thick line).}
\label{GMIs_M_16_EbN0}
\end{figure}

In \figref{GMIs_M_16_EbN0}, the MI and GMI for the three constellations under consideration are shown.\footnote{Calculated numerically via Monte Carlo integration.}  For PM-QPSK, the GMI and the MI coincide. This is not the case for the two other constellations. The results in \figref{GMIs_M_16_EbN0} show that $\Cfs$ gives a high MI at all SNRs; however, a large gap between the MI and GMI exists (more than $1$~dB for low code rates). Therefore, $\Cfs$ will not work well with a BW decoder. The situation is similar for SO-PM-QPSK, although in this case the losses are smaller. Interestingly, when comparing the GMIs for $\Cfs$ and SO-PM-QPSK, we observe that they cross at around $\R\approx 3.25$~bits/symbol. This indicates that a capacity-approaching transmitter with a BW decoder will perform better with $\Cfs$ than SO-PM-QPSK at high SNR. However, PM-QPSK is the best choice at any SNR.

\begin{figure}
\newcommand{\scale}{0.8}
\newcommand{\scalesmall}{0.55}
\centering
\psfrag{xlabel}[cc][cB][\scale]{SNR $\gamma$~[dB]}%
\psfrag{ylabel}[cc][cb][\scale]{$\BERout$}%
\psfrag{14}[cc][cc][\scale][-90]{{\fcolorbox{white}{white}{{$\Rc=1/4$}}}}%
\psfrag{12}[cc][cc][\scale][-90]{{\fcolorbox{white}{white}{{$\Rc=1/2$}}}}%
\psfrag{34}[cc][cc][\scale][-90]{{\fcolorbox{white}{white}{{$\Rc=3/4$}}}}%
\psfrag{91}[cc][cc][\scale][-90]{{\fcolorbox{white}{white}{{$\Rc=9/10$}}}}%
\includegraphics[width=\columnwidth]{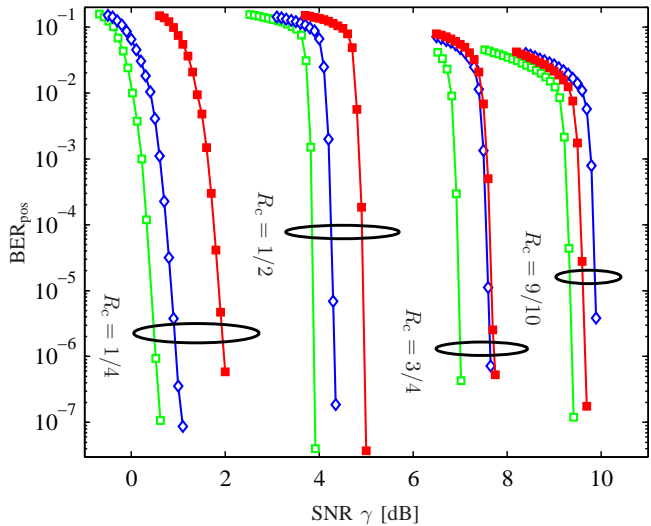}
\caption{Post-FEC BER ($\BERout$) for the LDPC code with different code rates and PM-QPSK (squares), SO-PM-QPSK (diamonds), and $\Cfs$ (filled squares).}
\label{BER_FER_EsN0}
\end{figure}

To show that the conclusions above correspond to gains in terms of $\BERout$, we consider the LDPC codes defined in \secref{Sec:NumResults:GMI} and the additional code rate $\Rc=1/4$ (also defined in \cite{ETSI_EN_302_307_v121}). The obtained BER results for $4$ different code rates are shown in \figref{BER_FER_EsN0}. Among the three constellations, PM-QPSK always gives the lowest $\BERout$. The gains offered by PM-QPSK with respect to $\Cfs$ for low code rates are about $1$~dB. More importantly, these gains are obtained by using a very simple demapper that computes four BPSK LLRs, one in each dimension. These results also show that the GMI curves in \figref{GMIs_M_16_EbN0} predict the coded performance of the system well. For example, the GMI curves indicate that at high code rates, $\Cfs$ is better than SO-PM-QPSK, which is exactly what happens in terms of $\BERout$ (\ie for $\Rc=9/10$, $\Cfs$ gives a lower $\BERout$ than SO-PM-QPSK).

\subsection{Achievable Rates for $M=256$}\label{M.256}

For $M=256$ (\ie $2$~bits/dimension) we consider two constellations. The first one is PM-16QAM, which is a straightforward generalization of PM-QPSK formed as the Cartesian product of four $4$-ary pulse amplitude modulation (PAM) constellations. The labeling problem for PM-16QAM then boils down to labeling a $4$-PAM constellation. Here we then consider the three nonequivalent binary labelings for $4$-PAM: the BRGC \cite{Gray53,Agrell07}, the natural binary code (NBC) \cite[Sec.~II-B]{Agrell10b}, and the anti-Gray code (AGC) \cite[Sec.~IV-E]{Alvarado12b}.

The second constellation we consider is a lattice-based constellation which we denote by $\Cft$. It consists of all points with integer coordinates, such that the coordinate sum is odd and the Euclidean norm is $3$ or less.\footnote{The same construction used with norm $1$ gives PS-QPSK.} In lattice terminology, $\Cft$ consists of the five first spherical shells of the $D_{4}$ lattice centered at a hole. The constellation was first characterized in \cite[Table.~IV]{Welti74} and \cite[p.~822]{Conway83} and it corresponds to a point on the solid line in \cite[Fig.~1~(a)]{Karlsson12OFC} ($4$~bits/symbol/pol).

To label this constellation, we use a numerically optimized labeling obtained using the binary-switching algorithm (BSA) and the GMI approximation in \cite{Alvarado14a}. The BSA was executed $300$ times, and every time initialized with a randomly generated seed. A labeling was obtained, optimized for an SNR of $\gamma=5$~dB (i.e., for MI around $3$~bits/symbol). Binary labelings that give a slightly higher GMI can be obtained when optimizing at lower SNR; however, the gains are marginal.

\begin{figure}[tbp]
\newcommand{\scale}{0.8}
\newcommand{\scalesmall}{0.65}
\centering
\psfrag{Sim.}[cl][cl][\scalesmall]{{\fcolorbox{white}{white}{{Sim. results at}}}}%
\psfrag{Res.}[cl][cl][\scalesmall]{{\fcolorbox{white}{white}{{$\BERout=10^{-4}$}}}}%
\psfrag{xlabel}[cc][cB][\scale]{$\Eb/\No$~[dB]}%
\psfrag{ylabel}[cc][cb][\scale]{$\R$~[bits/symbol]}%
\psfrag{CAW}[cl][cl][\scalesmall]{Channel Capacity}%
\psfrag{CapAWGNCapacityyyy}[cl][cl][\scalesmall]{Channel Capacity}%
\psfrag{MI}[cl][cl][\scalesmall]{MI}%
\psfrag{GMI}[cl][cl][\scalesmall]{GMI}%
\psfrag{SimPM16QAM}[cl][cl][\scalesmall]{PM-16QAM (BRGC)}%
\psfrag{SimC416}[cl][cl][\scalesmall]{$\Cfs$}%
\psfrag{PM-16QAM}[cl][cl][\scale]{PM-16QAM}%
\psfrag{PM-16QAM-BRGC}[cl][cl][\scale]{PM-16QAM (BRGC)}%
\psfrag{PM-16QAM-NBC}[cl][cl][\scale]{PM-16QAM (NBC)}%
\psfrag{PM-16QAM-AGC}[cl][cl][\scale]{PM-16QAM (AGC)}%
\psfrag{C4256}[cl][cl][\scale]{$\Cft$}%
\psfrag{C4256GMI}[cl][cl][\scalesmall]{$\Cft$ GMI}%
\includegraphics[width=\columnwidth]{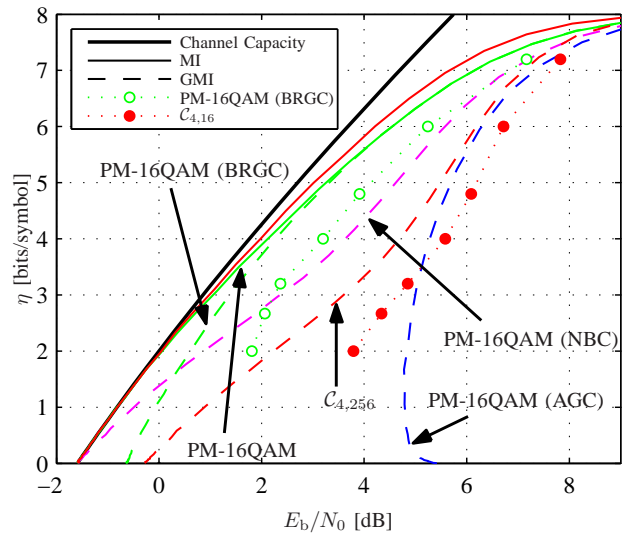}
\caption{MI and GMI for two constellations with $M=256$. Three labelings are considered for PM-16QAM. Simulations results are shown with markers for the seven code rates in \eqref{coding.rates} at $\BERout=10^{-4}$.}
\label{GMIs_M_256_EbN0}
\end{figure}

The obtained results are shown in \figref{GMIs_M_256_EbN0} and are quite similar to the ones in \figref{GMIs_M_16_EbN0}. When compared to PM-16QAM, the constellation $\Cft$ gives higher MI but lower GMI. We thus conclude that $\Cft$ is unsuitable for a BW decoder. A major advantage with PM-16QAM is the existence of a Gray code, which not only offers good performance but also lets the LLRs be calculated in each dimension separately, thus reducing complexity. The results in \figref{GMIs_M_256_EbN0} also show a quite large gap between the MIs for $\Cft$ and PM-16QAM in the high-SNR regime. This is explained by the increase in minimum Euclidean distance of $\Cft$ with respect to PM-16QAM \cite[Fig.~1~(a)]{Karlsson12OFC}.

To show that the performance of a BW decoder based on LDPC codes follows the GMI prediction, we simulated $7$ different code rates: $1/4$ and the ones in \eqref{coding.rates} (all of them defined in \cite{ETSI_EN_302_307_v121}), PM-16QAM labeled by the BRGC, and $\Cfs$ using the numerically optimized binary labeling. For each of the $14$ coding and modulation pairs, we measured the minimum value of $\Eb/\No$ needed to guarantee $\BERout=10^{-4}$. The obtained results are shown with circles in \figref{GMIs_M_256_EbN0}, where the vertical position of the marker is given by the achieved spectral efficiency (i.e., $\eta=\Rc m$). The obtained results clearly show that the BW decoder based on LDPC codes follow the GMI curve quite well. The SNR penalty of this particular family of LDPC codes with respect to the GMI is between $1$ and $0.5$~dB for low and high code rates, respectively.

\subsection{Achievable Rates for $M=4096$}\label{M.4096}

\begin{figure}[tbp]
\newcommand{\scale}{0.8}
\newcommand{\scalesmall}{0.65}
\centering
\psfrag{Sim.}[cl][cl][\scalesmall]{{\fcolorbox{white}{white}{{Sim. results at}}}}%
\psfrag{Res.}[cl][cl][\scalesmall]{{\fcolorbox{white}{white}{{$\BERout=10^{-4}$}}}}%
\psfrag{xlabel}[cc][cB][\scale]{$\Eb/\No$~[dB]}%
\psfrag{ylabel}[cc][cb][\scale]{$\R$~[bits/symbol]}%
\psfrag{CAW}[cl][cl][\scalesmall]{Channel Capacity}%
\psfrag{CapAWGNCapacityyyy}[cl][cl][\scalesmall]{Channel Capacity}%
\psfrag{MI}[cl][cl][\scalesmall]{MI}%
\psfrag{GMI}[cl][cl][\scalesmall]{GMI}%
\psfrag{SimPM64QAM}[cl][cl][\scalesmall]{PM-64QAM (BRGC)}%
\psfrag{SimC44096}[cl][cl][\scalesmall]{$\Cff$}%
\psfrag{PM64BRGCMI}[cl][cl][\scalesmall]{PM-64QAM MI}%
\psfrag{PM64BRGCGMIIIIIIII}[cl][cl][\scalesmall]{PM-64QAM (BRGC)}%
\psfrag{PM64NBCGMI}[cl][cl][\scalesmall]{PM-64QAM (NBC)}%
\psfrag{C44096MI}[cl][cl][\scalesmall]{$\Cff$ MI}%
\psfrag{C44096GMI Ebba}[cl][cl][\scalesmall]{$\Cff$ GMI}%
\psfrag{PM-64QAM}[cl][cl][\scale]{PM-64QAM}%
\psfrag{PM-64QAM-BRGC}[cl][cl][\scale]{PM-64QAM (BRGC)}%
\psfrag{PM-64QAM-NBC}[cl][cl][\scale]{PM-64QAM (NBC)}%
\psfrag{C44096}[cl][cl][\scale]{$\Cff$}%
\includegraphics[width=\columnwidth]{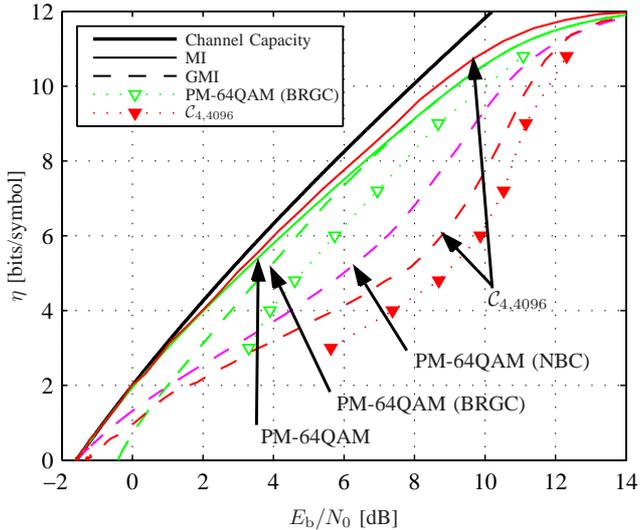}
\caption{MI and GMI for two constellations with $M=4096$. Simulations results are shown with markers for seven code rates at $\BERout=10^{-4}$.
}
\label{GMIs_M_4096_EbN0}
\end{figure}

For $M=4096$, we consider PM-64QAM labeled by the BRGC and by the NBC. This choice is motivated by the fact that the BRGC and the NBC are good labelings for the constituent $8$-PAM constellation in terms of GMI for high and low SNR, respectively. We also consider $\Cff$, which is the best known $4096$-point constellation for uncoded transmission at high SNR \cite{Agrell-codes}, it is a subset of the $D_4$ lattice, found by extensive numerical search, and its binary labeling was also numerically optimized.\footnote{The numerically optimized labelings obtained for $\Cfs$ and $\Cff$ have no regular structure and will be made available online as supplementary data to this paper.} The obtained results are shown in \figref{GMIs_M_4096_EbN0} and indicate that $4$D optimized constellation offer gains in terms of MI, however, when the GMI is considered, it performs suboptimally.\footnote{Due to the large number of constellation points and dimensions, the MI and GMI for $\Cft$ and $\Cff$ was estimated via Monte Carlo integration.} For example, at $\R=6$~bits/symbol, the losses caused by using $\Cff$ and a BW decoder with respect to PM-64QAM with the BRGC are about $4$~dB. Similarly to \figref{GMIs_M_256_EbN0}, \figref{GMIs_M_4096_EbN0} also shows the achieved spectral efficiencies for a target $\BERout=10^{-4}$ and the same code rates used in \secref{M.256}. The results show that the penalties caused by using $\Cff$ with respect to PM-64QAM are much larger than the corresponding penalties in \figref{GMIs_M_256_EbN0}.

For $M=4096$, the problem of selecting the binary labeling is very challenging. Although good labelings in the low- and high-SNR regimes can be found, these labelings are not necessarily suitable for the practically relevant medium-SNR regime. On the other hand, using $8$-PAM in each dimension simplifies the search for labelings and results in penalties (with respect to the MI) tending to zero for medium and high SNR values.

\begin{figure}[tbp]
\newcommand{\scale}{0.8}
\newcommand{\scalesmall}{0.65}
\centering
\psfrag{xlabel}[cc][cB][\scale]{$\Eb/\No$~[dB]}%
\psfrag{ylabel}[cc][cb][\scale]{$\R$~[bits/symbol]}%
\psfrag{CAW}[cl][cl][\scalesmall]{Channel Capacity}%
\psfrag{PMQGMI}[cl][cl][\scalesmall]{PM-QPSK GMI}%
\psfrag{C416MI}[cl][cl][\scalesmall]{$\Cfs$ MI}%
\psfrag{PM16BRGCGMIIIIIIIIII}[cl][cl][\scalesmall]{PM-16QAM (BRGC)}%
\psfrag{C4256MI}[cl][cl][\scalesmall]{$\Cft$ MI}%
\psfrag{PM64BRGCGMI}[cl][cl][\scalesmall]{PM-64QAM (BRGC)}%
\psfrag{C44096MI}[cl][cl][\scalesmall]{$\Cff$ MI}%
\psfrag{CapAWGNCapacity}[cl][cl][\scalesmall]{Channel Capacity}%
\psfrag{MI}[cl][cl][\scalesmall]{MI}%
\psfrag{GMI}[cl][cl][\scalesmall]{GMI}%
\includegraphics[width=\columnwidth]{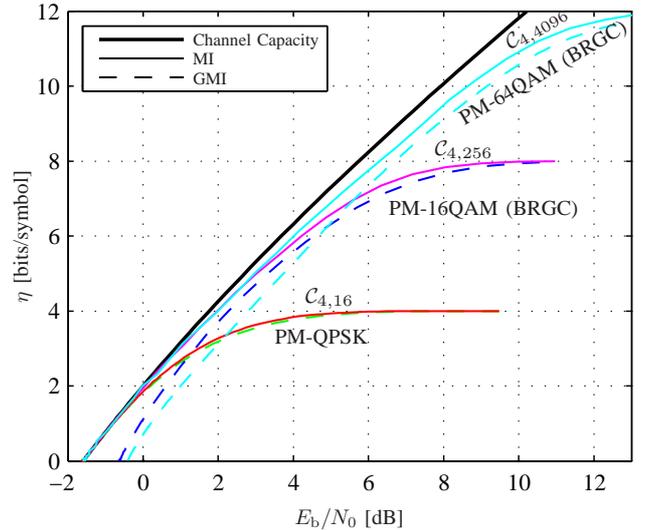}%
\overlay{-2.1cm}{6.5cm}{\rotatebox{30}{\footnotesize$\Cff$}}%
\overlay{-2.7cm}{5.5cm}{\rotatebox{30}{\footnotesize PM-64QAM (BRGC)}}%
\overlay{-3.0cm}{5.1cm}{\rotatebox{0}{\footnotesize $\Cft$}}%
\overlay{-3.6cm}{4.3cm}{\rotatebox{0}{\footnotesize PM-16QAM (BRGC)}}%
\overlay{-4.7cm}{3.14cm}{\rotatebox{0}{\footnotesize $\Cfs$}}%
\overlay{-5.1cm}{2.6cm}{\rotatebox{0}{\footnotesize PM-QPSK}}%
\caption{MI and GMI for the best constellations with $M=16$, $256$, and $4096$ from Figs.~\ref{GMIs_M_16_EbN0}, \ref{GMIs_M_256_EbN0}, and \ref{GMIs_M_4096_EbN0}, respectively.}
\label{MI_EbN0_all_together_MonteCarlo}
\end{figure}

To conclude, we selected the constellations and labelings that give the highest MI and GMIs in Figs.~\ref{GMIs_M_16_EbN0}, ~\ref{GMIs_M_256_EbN0}, and \ref{GMIs_M_4096_EbN0}. The results are presented in \figref{MI_EbN0_all_together_MonteCarlo} and show that the best constellation in terms of MI, regardless of the targeted spectral efficiency, is $\Cff$.\footnote{This is of course ignoring practical problems that would arise by using large constellations at low SNR.} The gap to the channel capacity for $\R\leq 10$~bits/symbol is less than $1$~dB, which makes us believe that changing the shape of a constellation with large cardinality is enough to make the MI to be close to the channel capacity. 

When the GMI is considered, the results in \figref{MI_EbN0_all_together_MonteCarlo} indicate that for $\R\leq 3$~bits/symbol, PM-QPSK should be the preferred alternative, for $3\leq \R \leq 6$~bits/symbol, PM-16QAM labeled by the BRGC should be used, and for $\R\geq6$~bits/symbol, PM-64QAM with the BRGC should be used. For $3\leq \R \leq 6$~bits/symbol and PM-16QAM, the optimum FEC overheads should then vary between $33.3$\% and $166$\%, which is good agreement with the code rates considered in \secref{Sec:NumResults:GMI} (see \eqref{coding.rates}). The results in this figure also show that for $\R\geq 3$~bits/symbol, the loss from using a BW decoder instead of an ML decoder is typically less than $1$~dB.

\section{Conclusions}\label{Sec:Conclusions}

In this paper, we studied achievable rates for coherent optical coded modulation transceivers where the receiver is based on a bit-wise structure. It was shown that the generalized mutual information is the correct metric to study the performance of capacity-approaching coded modulation transceivers based on this paradigm. We conjecture that the correct metric for a bit-wise receiver with iterative demapping is the mutual information.

For the suboptimal bit-wise structure under consideration, both analytical and numerical results show that simply transmitting and receiving independent data in each quadrature of each polarization is the best choice. Multidimensional constellations optimized for uncoded systems were shown to give high MI, and are thus good for ML decoders; these constellations, however, are not well-suited for bit-wise decoders. On top of the weaker performance and higher demapper complexity, such constellation also carry the design challenge of selecting a good binary labeling.

We did not try to increase the generalized mutual information by changing the shape of the constellation (geometrical shaping) or the probability of the transmitted symbols (probabilistic shaping). Constellation shaping and the effect of the nonlinear optical channel using the GMI as a figure of merit are left for future work. The intriguing connection between the generalized mutual information and the pre-FEC BER (see \figsref{BERout_vs_BERin}{BERout_vs_GMI}) is also left for further investigation.

\balance    
\section*{Acknowledgments}\label{Sec:Ack}

The authors would like to thank Dr. Domani\c{c} Lavery (University College London) and Tobias Fehenberger (Technische Universit\"{a}t M\"{u}nchen) for fruitful discussions regarding different parts of this manuscript.

\bibliography{IEEEabrv,references_all}
\bibliographystyle{IEEEtran}

\end{document}